\documentstyle[eqsecnum,aps,preprint]{revtex}
\begin{document}
\draft
\title{Branch Cuts due to Finite-Temperature Quasiparticles}
\author{H. Arthur Weldon}
\address{Department of Physics, West Virginia University, Morgantown WV
26506-6315}
\date{May 12, 1998}

\maketitle

\begin{abstract}

The branch points of individual thermal self-energy diagrams at $k^{2}=4m^{2}, k^{2}=9m^{2},\dots$
are shown not to be branch points of the full thermal self-energy.
Branch points of the full theory 
 are determined by the complex, temperature-dependent energies of the
quasiparticles, defined as  the pole location, $k_{0}={\cal E}(\vec{k})$, 
of the exact retarded propagator.  
The full retarded self-energy is found to have  branch points at
$k_{0}=2{\cal E}(\vec{k}/2)$ and
$k_{0}=3{\cal E}(\vec{k}/3)$ as well as cuts in the space-like region.
The discontinuities across the branch cuts are complex.
The advanced  self-energy is related by reflection to the retarded. 

\end{abstract}

\narrowtext

\section{Introduction}

At finite temperature, self-energy functions have more branch cuts and more complicated
discontinuities than at zero temperature. The finite-temperature discontinuities have direct
physical significance [1-9]. It is possible to compute the discontinuity of a self-energy
diagram without having to compute the real part by employing cutting rules that replace
certain propagators with Dirac delta functions [10-15].
All the known results about the location of branch cuts and the discontinuities across them
apply at each order of perturbation theory. The pertubation theory is defined by 
choosing free thermal propagators that have poles at  the zero-temperature mass $m$. This paper
will demonstrate that when perturbation theory is summed the full self-energy will have branch
cuts in different places and with different discontinuities than given in perturbation theory.

\subsection{ Example at Zero-Temperature}

A simple zero-temperature example for a scalar field with interaction  ${\cal
L}_{I}=g\phi^{3}/6$ will illustrate  how higher order corrections can shift the location of
branch cuts.  Suppose that  $m$ is the physical mass  but that
one performs pertubative calculations using a free propagator $\Delta(k)=1/[k^{2}-m_{0}^{2}]$,
where
$m_{0}$ is some different mass.  For simplicity $m_{0}$ should be finite and not the bare
mass.  The 
one-loop self-energy
\begin{equation}
\Pi^{(1)}(k)={ig^{2}\over 2}\int {d^{4}p\over(2\pi)^{4}}\Delta(p)\Delta(p-k)\end{equation}
 has a branch cut for $k^{2}\ge 4m_{0}^{2}$. The discontinuity across the branch cut is
\begin{eqnarray}
{\rm Disc}\,\Pi^{(1)}(k)=&&{-ig^{2}\over 8\pi^{2}}\!\int\! d^{4}p\,
\delta_{+}(p^{2}\!-\!m_{0}^{2})\delta_{+}((p-k)^{2}\!-\!m_{0}^{2})\nonumber\\
=&&{-ig^{2}\over 16\pi}\big(1-{4m_{0}^{2}\over k^{2}}\big)^{1/2}.
\end{eqnarray}
The indication that $k^{2}=4m_{0}^{2}$ is not a branch point of the full theory comes from
the two-loop contribution. The full propagator is $D'(k)=1/[k^{2}-m^{2}-\Pi(k)]$ and by 
definition $\Pi$ contains the necessary counter term to vanish at the true mass $k^{2}=m^{2}$.
To do pertubation theory with mass $m_{0}$ the full propagator is written
$D'(k)=1/[k^{2}-m_{0}^{2}-\tilde{\Pi}(k)]$ where 
$\tilde{\Pi}(k)=m^{2}-m_{0}^{2}+\tilde{\Pi}(k)$. Of course $\tilde{\Pi}$ does not vanish at
$k^{2}=m^{2}$ or at $k^{2}=m_{0}^{2}$ and this is the source of the problem.  
A self-energy insertion on the internal lines of (1.1) gives the two-loop self-energy term
\begin{equation}
\Pi^{(2)}(k)=ig^{2}\int{d^{4}p\over (2\pi)^{4}}\big[\Delta(p)\tilde{\Pi}(p)\Delta(p)\big]
\Delta(p-k).\end{equation}
This has a two-particle and a three-particle discontinuity. The quantity in square
brackets has a double pole at $p^{2}=m_{0}$.  Using
$[\Delta(p)]^{2}=-\partial\Delta(p)/\partial p^{2}$ lead to  a two-particle discontinuity
\begin{eqnarray}
{\rm Disc}\,\Pi^{(2)}(k)={ig^{2}\over 4\pi^{2}}\int d^{4}p&&\;
\delta_{+}'(p^{2}-m_{0}^{2})\,\tilde{\Pi}(p)\nonumber\\
&&\times\delta_{+}((p-k)^{2}-m_{0}^{2}).\end{eqnarray}
The presence of $\delta'(p^{2}-m_{0}^{2})$ requires an expansion of the shifted self-energy
for $p^{2}\approx m_{0}^{2}$:
\begin{displaymath}\tilde{\Pi}(p^{2})=\delta
m^{2}+\Pi(m_{0}^{2})+(p^{2}-m_{0}^{2})\Pi^{\prime}(m_{0}^{2})+\dots
\end{displaymath} where $\delta
m^{2}=m^{2}-m_{0}^{2}$. The integrated two-particle discontinuity is
\begin{displaymath}
{\rm Disc}\,\Pi^{(2)}(k)\!=\!{ig^{2}\over 8\pi}
\Big[{\delta m^{2}+\Pi(m_{0}^{2})\over
k^{2}\big(1-{4m_{0}^{2}\over k^{2}}\big)^{1/2}}
-\Pi^{\prime}(m_{0}^{2})\big(1-{4m_{0}^{2}\over k^{2}}\big)^{1/2}
\Big].
\end{displaymath}
The second term  changes the coefficient of (1.2) as required by wave function
renormalization. The first terms is more important: It
is infinite at $k^{2}=4m_{0}^{2}$. The infinity is a signal
that the correct branch point is not at $k^{2}=4m_{0}^{2}$.
Multiple  self-energy insertions on the same skeleton have two effects. First they modify the
coefficient of
$(1-4m_{0}^{2}/k^{2})^{-1/2}$ to be
\begin{eqnarray}
\delta
m^{2}+\Pi(m_{0}^{2})+&&\delta m^{2}\,\Pi^{\prime}(m_{0}^{2})+
{1\over 2}(\delta m^{2})^{2}\Pi^{\prime\prime}(m_{0}^{2})+\dots\nonumber\\
=&&\delta m^{2}+\Pi(m^{2})=\delta m^{2},\nonumber\end{eqnarray} where, in the last step, 
$\Pi(m^{2})=0$ has been used.   Thus only $\delta m^{2}$ survives as the coefficient of inverse
square root. Multiple  self-energy insertions  also produce successively higher powers of the
inverse square root:
\begin{eqnarray}
{\rm Disc}\,\Pi(k)=&&{-ig^{2}\over 8\pi}
\Big[\big(1-{4m_{0}^{2}\over k^{2}}\big)^{1/2}
-{2\delta m^{2}\over k^{2}}\big(1-{4m_{0}^{2}\over k^{2}}\big)^{-1/2}
\nonumber\\
&&\hskip0.5cm -{1\over 2}\big({2\delta m^{2}\over k^{2}}\big)^{2}
\big(1-{4m_{0}^{2}\over k^{2}}\big)^{-3/2}+\dots\Big].\nonumber\end{eqnarray}
This is the beginning of a Taylor series. All the corrections diverge at
the false threshold $k^{2}=4m_{0}^{2}$. In the range $4m_{0}^{2}<k^{2}<4(m_{0}^{2}+\delta
m^{2})$ each correction is finite but the Taylor series diverges. Thus pertubation theory
fails throughout this region of $k^{2}$.
To obtain a convergent series it is necessary to work in the range
$k^{2}>4(m_{0}^{2}+\delta m^{2})$. In this region the  Taylor series converges
 and  the sum is
the full two-particle discontinuity with branch point shifted to the physical mass
$m^{2}=m_{0}^{2}+\delta m^{2}$:
\begin{equation}
 {\rm Disc}\,\Pi(k)
={-ig^{2}\over 8\pi}\big(1-{4m^{2}\over k^{2}})^{1/2}.
\end{equation}
The true two-particle threshold is still a square root branch point at
$k^{2}=4m^{2}$. 
The breakdown of perturbation theory is entirely due to a propagator
$\Delta(k)=1/[k^{2}-m_{0}^{2}]$ with the the wrong mass $m_{0}$. The breakdown is easily
avoided by using $1/[k^{2}-m^{2}]$ for the free particle propagator.

\subsection{Non-zero Temperature}

In the previous example, individual diagrams of the perturbation series have branch points at
the wrong threshold $k^{2}=4m_{0}^{2}$ although the full theory does not. 
In finite-temperature field theory it is customary to perform 
perturbative calculations using free thermal propagators that have poles at
the zero-temperature, physical mass $m$. 
With this choice the one-loop self-energy has  a
 branch point at $k^{2}=4m^{2}$.  However this is not a true branch point of the
full theory. 
The insertion of the thermal self-energy 
 on an internal propagator produces a two-loop correction analogous to (1.3) in which there
is a double pole at $p^{2}=m^{2}$ because the one-loop self-energy  does not vanish 
there. The double pole produces a discontinuity proportional to $(1-m^{2}/k^{2})^{-1/2}$, which
diverges at $k^{2}=4m^{2}$.
This claim is easily checked by applying
 the Kobes-Semenoff cutting rules [10]  to compute  
the discontinuity. Both Le Bellac [14] and Gelis [15]    display the two-loop discontinuity as
an integral over $d^{4}p$ containing $\delta^{\prime} (p^{2}-m^{2})$.
This is the same structure as in (1.4). It is computed explicity in Appendix A and
 the result is proportional to
$(1-4m^{2}/k^{2})^{-1/2}$, which is infinite at the false threshold just as in the T=0 example.  

The branch points of the full theory are not obtained by trivially replacing $m^{2}$ by a
temperature-dependent effective mass. 
A proper calculation requires using  unperturbed propagators with poles at the
same energy at which the thermal self-energy vanishes so that there will be no double poles 
on internal lines.
An energy ${\cal E}$  which is a pole of the unperturbed propagator and also a zero of the
self-energy is automatically a pole in the full propagator.  
Poles in the full
propagator will occur at  energy $k_{0}={\cal E}(\vec{k})$ where ${\cal E}$ is
is an complicated function of $|\vec{k}|$ that depends on mass, coupling, and temperature. 
Moreover ${\cal E}$ is complex with  the imaginary part being the damping rate of the
 single particle excitation. For definiteness the real part of ${\cal E}$ will be chosen
positive and the imaginary part, negative. Thus ${\cal E}$ is in the fourth quadrant of the 
complex energy plane. The pole at  $k_{0}={\cal E}(\vec{k})$ is called the
quasiparticle pole. This paper will show that there is no branch cut at 
$k^{2}=4m^{2}$. Instead there is a two-quasiparticle branch point 
in the full self-energy at the complex, temperature-dependent energy $k_{0}=2{\cal
E}(\vec{k}/2)$.  The branch point is the end point of a branch cut in which the two
quasiparticles share the energy: $k_{0}={\cal E}(\vec{k}_{1})+{\cal E}(\vec{k}_{2})$ where
$\vec{k}=\vec{k}_{1}+\vec{k}_{2}$. 

The location of the quasiparticle pole in the full propagator  is determined
by effects that are higher order in the coupling. 
Approximating the full propagator by a simpler form that has a pole at the correct position 
reorders the perturbation series. This is similar to the Braaten-Pisarski resummmation [16,17]
of high temperature gauge theories but differs in several respects. First, the breakdown of 
perturbation theory near the false thresholds is not an infrared effect. The breakdown occurs
 even in theories with masses and even if the temperature is small. 
 Second, it is not necessary to retain the $k_{0}$ dependence of
the  self-energy in the new propagators, only the pole position 
${\cal E}(\vec{k})$.

A systematic method to organize  the reordering of perturbation theory 
is to employ the
integral equation  that relates the full self-energy to the exact Minkowski propagator
$D'_{ab}$ and  vertex $\Gamma$:
\begin{displaymath}
\Pi_{ad}(k)={ig^{2}\over 2}\!\int\!{d^{4}p\over
(2\pi)^{4}}D_{ab}^{\prime}(p)D_{ac}^{\prime}(p-k)\Gamma_{bcd}(p,p-k).
\end{displaymath}
Although one doesn't know the full propagator $D_{ab}^{\prime}$, 
the natural first approximation is to use a free quasiparticle propagator that
has a pole at the correct position. 
However the Minkowski integral equation is awkward to work with since it involves 
propagators with $2^{2}$ components and vertices with $2^{3}$ components. It is simpler to use
the imaginary-time formalism because there is  only one propagator and one vertex function.
In the imaginary-time approach the full self-energy is related to the full 
propagator
${\cal D}'$ and  vertex $\Gamma$ by the single integral equation
 \begin{eqnarray}
\Pi(\tau,\vec{k})={-g^{2}\over 2}\!\int\!
d\Omega_{12}\int_{0}^{\beta}&&\!d\tau'd\tau''\;
 {\cal D}'(\tau',\vec{k}_{1}){\cal D}'(\tau'',\vec{k}_{2})\nonumber\\
&&\times\Gamma(\tau',\vec{k}_{1};\tau'',\vec{k}_{2};\tau,\vec{k}), \end{eqnarray}
where
$d\Omega_{12}=d^{3}k_{1}d^{3}k_{2}\,\delta^{3}(\vec{k}-\vec{k}_{1}-\vec{k}_{2})/(2\pi)^{3}$. The
existence of a quasiparticle pole at $k_{0}={\cal E}(\vec{k})$ in the Minkowski propagator
determines the approximation to be used for the Euclidean propagator.
By Fourier transforming $\Pi(\tau,\vec{k})$ and then analytically continuing, it is possible to 
 obtain both the retarded and advanced self-energies $\Pi_{R/A}(k_{0},\vec{k})$.
This determines everything since
each of the four real-time propagators $D'_{ab}(k)$ are  linear combinations of the retarded and
advanced propagators.

The paper is organized as follows. Sec 2 discusses 
the exact retarded thermal propagator $D_{R}'(k)$ and separates the quasiparticle pole from
the self-energy effects. Sec 3 introduces the quasiparticle approximation to the propagator
in both Minkowski and Euclidean space-time.
Sec 4 computes the one-loop self-energy with quasiparticle propagators. The results are
displayed in (4.5) to (4.8). The calculations are performed in the imaginary time formalism
and then analytically continued to obtain $\Pi_{R/A}$. As a check, Appendix D performs the
one-loop calculation entirely in the real-time formalism. The calculation is more difficult
but produces exactly the same results. 
Sec 5 computes the two self-energy diagrams that contribute at
 two-loop order. The diagram in which there is a first-order self-energy insertion
on an internal line is the direct analogue of (1.3). Because the quasiparticle self-energy
vanishes at $k_{0}={\cal E}$, this diagram does not shift the location of the two
quasiparticle branch point. The effect of this contribution is a only a change in  the
coefficient of the two-quasiparticle cut. Both two-loop diagrams have
branch cuts for three-quasiparticle processes and these are computed.
Sec 6 contains the conclusions and the general relation between the real-time  $D_{ab}$ and
$\Pi_{ab}$ and the retarded/advanced quantities.

\section{Exact Propagators}

\subsection{Minkowski Space}

The quasiparticle poles occur in the Minkowski-space propagator
and it is necessary to begin there and then convert to Euclidean propagators.
The exact propagator in Minkowski-space has a
$2\times 2$ matrix structure. All four components 
are linear combinations of the exact retarded and advanced propagators $D_{R}'(k)$
and $D_{A}'(k)$ as displayed in (6.2).  Since $D_{A}'(k)=D_{R}'(-k)$ it suffices
to investigate $D_{R}'(k)$. The retarded propagator is analytic in the upper-half
of the complex $k_{0}$ plane and satisfies the condition
\begin{equation}
D'_{R}(k_{0},\vec{k})=
\big[ D'_{R}(-k_{0}^{*},\vec{k})\big]^{*}.\end{equation}
At zero temperature the exact propagator has poles at $k_{0}=\pm(m^{2}+\vec{k}^{2})^{1/2}$.
At non-zero temperature the location of these poles is temperature-dependent and complex.
For definiteness, let the
pole in the exact retarded propagator that occurs in the fourth quadrant be at
$k_{0}={\cal E}$ where \begin{equation}
{\cal E}(\vec{k})=E(\vec{k})-i\Gamma(\vec{k})/2\hskip1cm
(E>0; \Gamma>0).\end{equation}
Both $E$ and $\Gamma$ are complicated functions of momentum, temperature, and coupling.
The complex energy ${\cal E}$ will be called the quasiparticle energy. 
Because of (2.1) the retarded propagator must also have a pole in the third quadrant at
$k_{0}=-{\cal E}^{*}$. Also because of (2.1) the residues of these two poles are related:
\begin{eqnarray}\lim_{k_{0}\to{\cal E}}\;(k_{0}-{\cal E})D_{R}'(k)=&& Z/2E\nonumber\\
\lim_{k_{0}\to -{\cal E}^{*}}\;(k_{0}+{\cal E}^{*})D_{R}'(k)=
&&-Z^{*}/2E.\end{eqnarray} Here $Z$ plays the role of the wave-function renormalization constant. 
The retarded propagator is directly related to the retarded self-energy
\begin{equation}
D'_{R}(k)=\Big[k^{2}-m^{2}-\Pi_{R}(k)\Big]^{-1}.
\end{equation}
Because the full retarded propagator does not have 
poles at $k^{2}=m^{2}$, 
the proper self-energy  $\Pi_{R}(k)$ does not vanish at $k^{2}=m^{2}$. Therefore the
 usual Dyson-Schwinger expansion 
\begin{equation}
{1\over k^{2}-m^{2}}+{\Pi_{R}(k)\over (k^{2}-m^{2})^{2}}
+{\Pi_{R}^{2}(k)\over (k^{2}-m^{2})^{3}}+\cdots\end{equation}
is not useful. The first term has a simple pole at $k^{2}=m^{2}$, the second term
has a double pole, the third term has a triple pole,\dots. Performing
perturbation theory around $k^{2}=m^{2}$ is quite misleading. It is much better
to write the full retarded propagator as
\begin{equation}
D'_{R}(k)=\Big[(k_{0}-{\cal E})(k_{0}+{\cal
E}^{*})-\Pi_{Rqp}(k)\Big]^{-1}, \end{equation}
where the retarded quasiparticle self-energy is defined by
\begin{equation}
\Pi_{Rqp}(k)=\Pi_{R}(k)+\vec{k}^{2}+m^{2}-|{\cal E}|^{2}+i\Gamma k_{0}.
\end{equation}
By construction $\Pi_{Rqp}(k)$ vanishes at $k_{0}={\cal E}$ and also at $k_{0}=-{\cal
E}^{*}$:
\begin{equation}\Pi_{Rqp}({\cal E})=0\hskip1cm
\Pi_{Rqp}(-{\cal E}^{*})=0.\end{equation}
 The natural expansion around the quasiparticle poles  is
\begin{equation}
{1\over (k_{0}-{\cal E})(k_{0}+{\cal E}^{*})}
+{\Pi_{Rqp}(k)\over (k_{0}-{\cal E})^{2}(k_{0}+{\cal E}^{*})^{2}}
+\dots\end{equation}
The second term has only simple poles at $k_{0}={\cal E}$ and at
$k_{0}=-{\cal E}^{*}$.
It is convenient to define the derivative of the self-energy at these positions in
terms of a complex constant $B$:
\begin{equation}
{d\Pi_{Rqp}(k_{0})\over dk_{0}}
=\cases{2E\,B & $k_{0}={\cal E}$\cr\cr
-2E\,B^{*} & $k_{0}=-{\cal E}^{*}$}.\end{equation}
This constant $B$ is related to  the wave-function renormalization constant in (2.3) by
\begin{equation}
Z=1+B+B^{2}+\dots=1/(1-B).\end{equation}

It will be helpful to have similar results for the advanced propagator.
From the definition
 \begin{equation}
D_{A}'(k)=D_{R}'(-k),\end{equation}
the advanced propagator is analytic in the lower-half of the complex $k_{0}$ plane.
It must have poles in upper-half plane at $k_{0}={\cal E}^{*}$ and $k_{0}=-{\cal E}$.
To emphasize these poles it is convenient to write the advanced propagator as
\begin{equation}
D'_{A}(k)=\Big[(k_{0}+{\cal E})(k_{0}-{\cal
E}^{*})-\Pi_{Aqp}(k)\Big]^{-1}, \end{equation}
where the advanced quasiparticle self-energy is defined to be
\begin{equation}
\Pi_{Aqp}(k)=\Pi_{A}(k)+\vec{k}^{2}+m^{2}-|{\cal E}|^{2}-i\Gamma k_{0}.
\end{equation}

\subsection{Euclidean Space}

The finite-temperature Euclidean propagator is defined at discrete, imaginary frequencies
\begin{displaymath}\omega_{n}=i2\pi nT,\end{displaymath}
where $n$ is any integer. The full Euclidean propagator is
\begin{equation}
{\cal D}'(i\omega_{n},\vec{k})=\cases{-D'_{R}(i\omega_{n},\vec{k})
& if $n\ge 0$\cr 
-D'_{A}(i\omega_{n},\vec{k}) & if $n\le 0$},\end{equation}
with the overall minus sign chosen for later convenience. 
Relation (2.12) for $k_{0}$ imaginary implies  that 
$D'_{R}(i2\pi|n|T,\vec{k})=D'_{A}(-i2\pi|n|T,\vec{k})$. It follows that
${\cal D}'(i\omega_{n},\vec{k})$ is an even function of $n$.
The Euclidean propagator may be expressed in  terms of the self-energy as
\begin{displaymath}
{\cal D}'(i\omega_{n},\vec{k})=\Big[(\omega_{n})^{2}+\vec{k}^{2}+m^{2}
+\Pi(i\omega_{n},\vec{k})\Big]^{-1}\end{displaymath}
where the Euclidean self-energy is
\begin{equation}
\Pi(i\omega_{n},\vec{k})=\cases{\Pi_{R}(i\omega_{n},\vec{k})
& if $n\ge 0$\cr 
\Pi_{A}(i\omega_{n},\vec{k}) & if $n\le 0$}.\end{equation}
The relations
$\Pi_{R}(k_{0},\vec{k})=[\Pi_{R}(-k_{0}^{*},\vec{k})]^{*}$
and $\Pi_{A}(k_{0},\vec{k})=[\Pi_{A}(-k_{0}^{*},\vec{k})]^{*}$ guarantee that 
 (2.16) is real. 
To emphasize the quasiparticle aspect the propagator may be written
\begin{equation}{\cal D}'(i\omega_{n},\vec{k})=
\Big[-(i|\omega_{n}|-{\cal E})(i|\omega_{n}|+{\cal E}^{*})+\Pi_{qp}
(i\omega_{n},\vec{k})\Big]^{-1}\end{equation}
where the quasiparticle self-energy is
\begin{equation}
\Pi_{qp}(i\omega_{n},\vec{k})=\Pi(i\omega_{n},\vec{k})
+\vec{k}^{2}+m^{2}-|{\cal E}|^{2}-\Gamma|\omega_{n}|.\end{equation}
The presence of $|n|$ rather that $n$ in these results is very important but will
cause complications later.

\section{Quasiparticle Propagator}

The natural approximation to the full
Minkowski-space propagators is to retain the quasiparticle poles. Thus
approximate (2.6) and (2.13) by
\begin{eqnarray}
 D_{R}(k)=&&{1\over (k_{0}-{\cal E})(k_{0}+{\cal E}^{*})}\nonumber\\
D_{A}(k)=&&{1\over (k_{0}+{\cal E})(k_{0}-{\cal E}^{*})}.\end{eqnarray}
The corresponding Euclidean propagator for free quasiparticles follows from (2.17):
\begin{eqnarray}
{\cal D}(i\omega_{n},\vec{k})=&&{-1\over (i|\omega_{n}|-{\cal
E})(i|\omega_{n}|+{\cal E}^{*})}\nonumber\\
=&&{1\over \omega_{n}^{2}+\Gamma|\omega_{n}|
+|{\cal E}|^{2}}.\end{eqnarray}
The transform to  Euclidean time requires the Fourier summation
\begin{equation}
{\cal D}(\tau,\vec{k})=T\sum_{n=-\infty}^{\infty}e^{-i\omega_{n}\tau}\,
{\cal D}(i\omega_{n},\vec{k})\end{equation}
for $-\beta\le\tau\le\beta$.
 Since (3.2) is an even function of the integer $n$, (3.3) is automatically an even function of
$\tau$. 
To perform the summation   it is convenient 
to write (3.2)  without the absolute value bars on $n$ as
\begin{displaymath}
{-1\over (i\omega_{n}-{\cal
E})(i\omega_{n}+{\cal E}^{*})}
-\theta(-n){2i\omega_{n}({\cal E}^{*}-{\cal E})\over
(\omega_{n}^{2}+{\cal E}^{2})(\omega_{n}^{2}+{\cal
E}^{*2})}.\end{displaymath} Using this gives for the Fourier sum
\begin{eqnarray}
{\cal D}(\tau,\vec{k})=&&{1\over 2E}\Big([1+n({\cal E})]e^{-{\cal E}|\tau|}
+n({\cal E}^{*})e^{{\cal E}^{*}|\tau|}\Big)\nonumber\\
-&&T\sum_{n=1}^{\infty}e^{i\omega_{n} |\tau|}{2\Gamma\omega_{n}\over
(\omega_{n}^{2}+{\cal E}^{2})(\omega_{n}^{2}+{\cal E}^{*2})}.\end{eqnarray}
This is the form of the quasiparticle propagator that will be used in the subsequent self-energy
calculations. All the $\tau$ dependence   is of the form
$\exp(-\Lambda|\tau|)$ where $\Lambda$ is a member  of the set below
\begin{equation}
\Lambda\in\{ {\cal E}, -{\cal E}^{*}, -i\omega_{1}, -i\omega_{2}, -i\omega_{3},
\dots\}\hskip0.5cm {\rm Im}\Lambda<0.\end{equation}
Each $\Lambda$ has a negative imaginary part. The propagator will be written compactly as
\begin{equation}
{\cal D}(\tau)=\sum_{\Lambda}f(\Lambda)e^{-\Lambda|\tau|}\end{equation} 
in which the coefficient  functions are
\begin{eqnarray}
f({\cal E})=&&[1+n({\cal E})]/2E\nonumber\\
f(-{\cal E}^{*})=&&n({\cal E}^{*})/2E\\
f(-i\omega_{\ell})=&&-2T\Gamma\omega_{\ell}/ 
(\omega_{\ell}^{2}+{\cal E}^{2})(\omega_{\ell}^{2}+{\cal
E}^{*2}). \nonumber\end{eqnarray}

Although (3.4) will be used throughout, the infinite sum conceals several properties
that are important to note. First, the time dependence $\exp(-\Lambda|\tau|)$ with
Im $\Lambda<0$ will lead to a Euclidean self-energy that can be easily extended to
the retarded self-energy in Minkowski space. However the starting point ${\cal
D}(i\omega_{n},\vec{k})$ in (3.2) favors neither the retarded nor the advanced forms. 
Although it is not apparent, (3.4) is actually real:
\begin{equation}
{\cal D}(\tau,\vec{k})^{*}={\cal D}(\tau,\vec{k}).\end{equation}
This allows the time dependence to also be written $\exp(-\Lambda^{*}|\tau|)$
if continuation to the advanced form of the Minkowski self-energy is desired. 
Although (3.8) is not obvious, it must be true since ${\cal D}(i\omega_{n},\vec{k})$ is real and 
an even function of $n$. Appendix B proves (3.8) explicitly. 
Second, since $\exp(i\omega_{n}\beta)=1$ the quasiparticle propagator (3.3) satisfies the KMS
condition
\begin{equation}
{\cal D}(\beta-\tau,\vec{k})={\cal D}(\tau,\vec{k}).\end{equation} 
Without the infinite sum in (3.4) the KMS property would not hold. 
Appendix B proves (3.9) explicitly. 
Third, another way to obtain (3.4) is to begin with the time-ordered propagator in Minkowski
space, which is given by the following linear combination of the retarded and advanced propagators:
\begin{equation}
D_{11}(k)={[1+n(k_{0})]\over (k_{0}-{\cal E})(k_{0}+{\cal E}^{*})}
-{n(k_{0})\over (k_{0}+{\cal E})(k_{0}-{\cal E}^{*})}.\end{equation}
The Fourier transform, $D_{11}(t,\vec{k})$, for real positive time $t$
is determined by all the poles in the lower-half of the complex $k_{0}$ plane.
These poles are at $k_{0}={\cal E}, k_{0}=-{\cal E}^{*}$ and at
$k_{0}=-i\omega_{n}$ for $n>0$. 
The propagator in Euclidean time results from continuing from positive, real $t$
to negative, imaginary time $-i\tau$. The Euclidean propagator is ${\cal
D}(\tau,\vec{k})=iD_{11}(-i\tau,\vec{k})$ and gives precisely (3.4).

\section{One-Loop Self-Energy}

It is always easy to perform loop corrections by integrating over Euclidean time
and then Fourier transforming [1,18]. That method will be employed here. 
The first approximation to the integral equation (1.6) for the full self-energy
is to use the quasiparticle propagator (3.4) and the bare vertex without corrections.
This approximation treats the energy ${\cal E}$ exactly even though it is a function
of the coupling $g$. The one-loop correction shown in Fig. 1 is
\begin{equation}
\Pi^{I}(\tau,\vec{k})={-g^{2}\over 2}\int d\Omega_{12}\;
{\cal D}(\tau,\vec{k}_{1}){\cal D}(\tau,\vec{k}_{2}).
 \end{equation}
This may be expressed concisely  
using the notation (3.6) for the propagators:
\begin{equation}
\Pi^{I}(\tau,\vec{k})={-g^{2}\over 2}\int d\Omega_{12}\;
\sum_{\Lambda_{1},\Lambda_{2}}
f(\Lambda_{1})f(\Lambda_{2})e^{-(\Lambda_{1}+\Lambda_{2})|\tau|}.
 \end{equation}
The transform from $\tau$ to discrete frequency $\omega_{n}$ is
\begin{eqnarray}
\Pi^{I}(i\omega_{n},\vec{k})=&&\int_{0}^{\beta}d\tau
e^{i\omega_{n}\tau}\Pi^{I}(\tau,\vec{k})\\
=&&{g^{2}\over 2}\int d\Omega_{12}\;
\sum_{\Lambda_{1},\Lambda_{2}}
f(\Lambda_{1})f(\Lambda_{2}){1-e^{-(\Lambda_{1}+\Lambda_{2})\beta}
\over i\omega_{n}-\Lambda_{1}-\Lambda_{2}}.
 \nonumber\end{eqnarray}
This can be  extended  from  $i\omega_{n}$ for $n>0$ to complex $k_{0}$ with Im $k_{0}>0$.
It is analytic for Im $k_{0}>0$ because $\Lambda_{1}$ and $\Lambda_{2}$ have negative imaginary
parts. The extension therefore gives the retarded self-energy:
\begin{equation}
\Pi_{R}^{I}(k_{0},\vec{k})
={g^{2}\over 2}\int\! d\Omega_{12}\!
\sum_{\Lambda}
f(\Lambda_{1})f(\Lambda_{2}){1-e^{-(\Lambda_{1}+\Lambda_{2})\beta}
\over k_{0}-\Lambda_{1}-\Lambda_{2}}.
\end{equation}
Although this is analytic for $k_{0}$ in the upper-half of the complex plane, 
 when $k_{0}$ is
continued into the lower half-plane, the singularities at $k_{0}=\Lambda_{1}+\Lambda_{2}$
produce  branch cuts in the self-energy. 

{\it Physical Cuts:} It is useful to write out the various
cases for the different $\Lambda_{i}$. First, if $\Lambda_{1}={\cal E}_{1}$ and
$\Lambda_{2}={\cal E}_{2}$ the contribution to the self-energy is
\begin{equation}
{g^{2}\over 2}\int{d\Omega_{12}\over 2E_{1}2E_{2}}
{[1+n({\cal E}_{1})][1+n({\cal E}_{2})]-n({\cal E}_{1})n({\cal E}_{2})\over k_{0}-{\cal
E}_{1}-{\cal E}_{2}}.
\end{equation}
The discontinuity across the cut is complex. The statistical factors provide for the
Bose-Einstein enhanced emission of two quasiparticles minus the absorption of two
quasiparticles.   The second contribution is for
$\Lambda_{1}={\cal E}_{1}$ and
$\Lambda_{2}=-{\cal E}_{2}^{*}$:
\begin{equation}
{g^{2}\over 2}\int{d\Omega_{12}\over 2E_{1}2E_{2}}
{[1+n({\cal E}_{1})]n({\cal E}_{2}^{*})-n({\cal E}_{1})[1+n({\cal E}_{2}^{*})]\over k_{0}-{\cal
E}_{1}+{\cal E}_{2}^{*}}.
\end{equation}
The statistical factors account for a direct process in which quasiparticle 1 is emitted and
quasiparticle 2 is absorbed minus the inverse process.
If $\Lambda_{1}=-{\cal E}_{1}^{*}$ and
$\Lambda_{2}={\cal E}_{2}$ the result is
\begin{equation}
{g^{2}\over 2}\int{d\Omega_{12}\over 2E_{1}2E_{2}}
{n({\cal E}_{1}^{*})[1+n({\cal E}_{2})]-[1+n({\cal E}_{1}^{*})]n({\cal E}_{2})\over
k_{0}+{\cal E}_{1}^{*}-{\cal E}_{2}}.
\end{equation}
If  $\Lambda_{1}=-{\cal E}_{1}^{*}$ and
$\Lambda_{2}=-{\cal E}_{2}^{*}$ the self-energy is
\begin{equation}
{g^{2}\over 2}\int{d\Omega_{12}\over 2E_{1}2E_{2}}
{n({\cal E}_{1}^{*})n({\cal E}_{2}^{*})-[1+n({\cal E}_{1}^{*})]
[1+n({\cal E}_{2}^{*})]\over k_{0}+{\cal
E}_{1}^{*}+{\cal E}_{2}^{*}}.
\end{equation}
The one-loop self-energy is the sum  of (4.5)-(4.8) and (4.9) displayed below.
 Appendix D computes the same quantity using the Minkowski propagators and obtains 
exactly the same answer. 

{\it Unphysical Cuts:} There are some additional contributions to (4.3).
If both $\Lambda_{1}$ and $\Lambda_{2}$ are positive integer multiples of $-i2\pi T$ then
the numerator of (4.3) vanishes.
 However if only one of the $\Lambda_{j}$ is a  positive integer multiple of $-i2\pi T$
the numerator does not vanish.
Since (4.3) is symmetric under interchange of $\vec{k}_{1}\leftrightarrow\vec{k}_{2}$
it is only necessary to consider the case $\Lambda_{1}=-i\omega_{\ell}$,
$\Lambda_{2}={\cal E}_{2}$ or $-{\cal E}_{2}^{*}$ and double the result to obtain 
\begin{eqnarray}g^{2}\int {d\Omega_{12}\over 2E_{2}}
&&\sum_{\ell=1}^{\infty}{2T\,\Gamma_{1}\omega_{\ell}\over (\omega_{\ell}^{2}
+{\cal E}_{1}^{2})(\omega_{\ell}^{2}+{\cal E}_{1}^{*2})}\nonumber\\
&&\times\Big[{-1\over k_{0}+i\omega_{\ell}-{\cal E}_{2}}
+{1\over k_{0}+i\omega_{\ell}+{\cal E}_{2}^{*}}\Big].\end{eqnarray}
These terms have branch cuts in the lower half-plane  at $k_{0}=-i\omega_{\ell}+{\cal E}_{2}$ and
at $k_{0}=-i\omega_{\ell}-{\cal E}_{2}^{*}$. The cuts are unphysical in that they are
not entirely due to quasiparticle thresholds.  The
coefficient of this cut is proportional to the damping rate $\Gamma_{1}$ and is in this sense a
higher order effect. Sec 5 will show that (4.9) is exactly canceled by two-loop effects. 
For later comparison it is useful to 
return to the term in $\Pi^{I}(\tau,\vec{k})$ whose frequency transform produced this cut:
\begin{eqnarray}
g^{2}\int d\Omega_{12} &&
\sum_{\ell=1}^{\infty}e^{i\omega_{\ell}\tau}
{2T\,\Gamma_{1}\omega_{\ell}
\over (\omega_{\ell}^{2}+{\cal E}_{1}^{2})(\omega_{\ell}^{2}+{\cal E}_{1}^{*2})}
\nonumber\\
\times&&\Big[[1+n({\cal E}_{2})]{1\over 2E_{2}}e^{-{\cal E}_{2}\tau}
+n({\cal E}^{*}_{2}){1\over 2E_{2}}e^{{\cal E}^{*}_{2}\tau}\Big].
\end{eqnarray}

{\it Advanced Self-Energy:} Since quasiparticle propagator satisfies the KMS condition, the
integrand of (4.1) could equally be written
${\cal D}(\beta-\tau,\vec{k}_{1}){\cal D}(\beta-\tau,\vec{k}_{2})$. The
Fourier transform to $i\omega_{n}$  is then expressed as
\begin{displaymath}
\Pi^{I}(i\omega_{n},\vec{k})
={g^{2}\over 2}\int d\Omega_{12}\;
\sum_{\Lambda_{1},\Lambda_{2}}
f(\Lambda_{1})(\Lambda_{2}){e^{-(\Lambda_{1}+\Lambda_{2})\beta}-1
\over i\omega_{n}+\Lambda_{1}+\Lambda_{2}}.
\end{displaymath}
This is exactly the same self-energy as (4.3). However in this form it is easily extended from 
$i\omega_{n}$ to a function of complex $k_{0}$ that is analytic for Im $k_{0}<0$. This extension
 gives the advanced self-energy
\begin{displaymath}
\Pi^{I}_{A}(k_{0},\vec{k})
={g^{2}\over 2}\int d\Omega_{12}\;
\sum_{\Lambda_{1},\Lambda_{2}}
f(\Lambda_{1})f(\Lambda_{2}){e^{-(\Lambda_{1}+\Lambda_{2})\beta}-1
\over k_{0}+\Lambda_{1}+\Lambda_{2}}.
\end{displaymath}
It satisfies $\Pi_{A}(k)=\Pi_{R}(-k)$ as required and has all its branch points in the upper
half of the complex $k_{0}$ plane.

{\it Mixed Representations:} Because of the KMS condition one can also represent
the self-energy using a mixed form ${\cal D}(\tau,\vec{k}_{1}){\cal
D}(\beta-\tau,\vec{k}_{2})$. This leads to
\begin{displaymath}
\Pi^{I}(i\omega_{n},\vec{k})
={g^{2}\over 2}\int d\Omega_{12}\;
\sum_{\Lambda_{1},\Lambda_{2}}
f(\Lambda_{1})(\Lambda_{2}){e^{-\Lambda_{2}\beta}-e^{-\Lambda_{1}\beta}
\over i\omega_{n}-\Lambda_{1}+\Lambda_{2}}.
\end{displaymath}
Although this is the same self-energy, this representation cannot be 
easily extended to either the retarded or the advanced form of the self-energy.
In Sec 5B it will be necessary to use the KMS identity in a similar way to manipulate
the two-loop self-energy into a form whose Fourier transform will be  analytic
in the lower half-plane.

\section{Two-Loop Self-Energy}

The simplicity of the one-loop calculation makes it likely that the two-loop contributions 
 can be computed by the same method.  
The  contributions of 
Figs. 2 and 3   will be denoted by $\Pi^{II}_{A}$ and $\Pi^{II}_{B}$ respectively.

\subsection{Self-Energy Insertion on Quasiparticle Propagator}

The value of the diagram shown in Fig. 2  is
\begin{equation}
\Pi^{II}_{A}(\tau,\vec{k})=-g^{2}\int d\Omega_{12}\;
{\cal  D}^{I}(\tau,\vec{k}_{1}){\cal D}_{2}(\tau,\vec{k}_{2}),\end{equation}
where ${\cal D}^{I}$ is the one-loop corrected propagator:
\begin{displaymath}{\cal D}^{I}(\tau,\vec{k})
=\!\int_{0}^{\beta}\!d\tau'd\tau''
{\cal D}(\tau-\tau'',\vec{k})\Pi_{qp}(\tau''-\tau',\vec{k}){\cal
D}(\tau',\vec{k}).\end{displaymath} This is not the most convenient way to compute ${\cal
D}^{I}$. It is easier to employ the method discussed after (3.10).  This requires the
Minkowski-space time-ordered propagator, now with one  insertion of the retarded and
advanced self-energies:
 \begin{displaymath}D_{11}^{I}(k)=
{[1+n(k_{0})]\Pi_{Rqp}^{I}(k)\over (k_{0}-{\cal E})^{2}(k_{0}+{\cal E}^{*})^{2}}
-{n(k_{0})\Pi_{Aqp}^{I}(k)\over (k_{0}+{\cal E})^{2}(k_{0}-{\cal E}^{*})^{2}}.
\end{displaymath}
To Fourier transform to real, positive time $t$ requires closing the $k_{0}$
contour in the lower-half of the complex $k_{0}$ plane. The singularities in
$k_{0}$ in the lower-half plane are as follows: (1) a simple pole at $k_{0}={\cal
E}$, (2) a simple pole at $k_{0}=-{\cal E}^{*}$, (3) simple poles in $n(k_{0})$ at
$k_{0}=-i\omega_{\ell}$, and (4) branch cuts in $\Pi_{Rqp}(k)$. 
Consequently the Fourier transform is
\begin{eqnarray}
iD_{11}^{I}(t,\vec{k})=&&
[1+n({\cal E})]{B\over 2E}e^{-i{\cal E}t}
+n({\cal E}^{*}){B^{*}\over 2E}e^{i{\cal E}^{*}t}\nonumber\\
+&&T\sum_{\ell=1}^{\infty}\Big[
{\Pi_{Rqp}^{I}(k)e^{-ik_{0}t}\over (k_{0}-{\cal E})^{2}(k_{0}+{\cal E}^{*})^{2}}
\Big]_{k_{0}=-i\omega_{\ell}}\nonumber\\
-&&T\sum_{\ell=1}^{\infty}\Big[
{\Pi_{Aqp}^{I}(k)e^{-ik_{0}t}\over (k_{0}+{\cal E})^{2}(k_{0}-{\cal E}^{*})^{2}}
\Big]_{k_{0}=-i\omega_{\ell}}\nonumber\\
&&+i\int_{Im\;k_{0}<0}^{\rm cuts}{dk_{0}\over 2\pi}D_{11}^{I}(k_{0})e^{-ik_{0}t}.
 \nonumber\end{eqnarray}
The self-energies $\Pi_{Rqp}$ and $\Pi_{Aqp}$ can be expressed in terms of
$\Pi_{R}$ and $\Pi_{A}$ using the definitions (2.7) and (2.14). Evaluating the propagator at the
Euclidean time
$t=-i\tau$ gives
\begin{eqnarray}
{\cal D}^{I}(\tau,\vec{k})=&&
[1+n({\cal E})]{B\over 2E}e^{-{\cal E}\tau}
+n({\cal E}^{*}){B^{*}\over 2E}e^{{\cal E}^{*}\tau}\nonumber\\
+&&T\sum_{\ell=1}^{\infty}e^{i\omega_{\ell}\tau}\;
\Big[{2\Gamma\omega_{\ell}
\over (\omega_{\ell}^{2}+{\cal E}^{2})(\omega_{\ell}^{2}+{\cal E}^{*2})}
\nonumber\\
+&&{4\Gamma\omega_{\ell}\big\{(\omega_{\ell}^{2}+{\cal E}
{\cal E}^{*})(\vec{k}^{2}+m^{2}-{\cal E}{\cal E}^{*})+\Gamma^{2}\omega_{\ell}^{2}
\big\}\over(\omega_{\ell}^{2}+{\cal E}^{2})^{2}(\omega_{\ell}^{2}
+{\cal E}^{*2})^{2}}\nonumber\\
+&&{\Pi_{R}^{I}(-i\omega_{\ell})\over (\omega_{\ell}^{2}+{\cal E}{\cal E}^{*}
-\Gamma\omega_{\ell})^{2}}
-{\Pi_{R}^{I}(i\omega_{\ell})\over (\omega_{\ell}^{2}+{\cal E}{\cal E}^{*}
+\Gamma\omega_{\ell})^{2}}\Big]\nonumber\\
&&+i\int_{Im\;k_{0}<0}^{\rm cuts}{dk_{0}\over 2\pi}D_{11}^{I}(k_{0})e^{-k_{0}\tau}.
\end{eqnarray}
One way of proceeding is to add this correction to the free quasiparticle propagator (3.4).
In the sum ${\cal D}+{\cal D}^{I}$ the coefficients of the quasiparticle terms are
modified to $1+B$ and $1+B^{*}$ and the term proportional to $\Gamma$ in (3.4) cancels
in the sum. It was this term that produced the unphysical cuts in the one-loop self-energy.
The cancellation in  ${\cal D}+{\cal D}^{I}$  guarantees that  unphysical one-loop  cuts will
be canceled in two-loop order. The following discussion shows these features explicitly as
well as the three-quasiparticle cuts that arise.

{\it Wave Function Correction to the Two Quasiparticle Cut:}
To compute the self-energy requires substituting (5.2)  into (5.1).
The contribution of the first line of (5.2) to $\Pi^{II}_{A}(\tau,\vec{k})$ is  
\begin{eqnarray}
-g^{2}\int d\Omega_{12} &&\Big[
[1+n({\cal E}_{1})]{B_{1}\over 2E_{1}}e^{-{\cal E}_{1}\tau}
+n({\cal E}^{*}_{1}){B^{*}_{1}\over 2E_{1}}e^{{\cal
E}^{*}_{1}\tau}\Big]\nonumber\\
\times&&\Big[[1+n({\cal E}_{2})]{1\over 2E_{2}}e^{-{\cal E}_{2}\tau}
+n({\cal E}^{*}_{2}){1\over 2E_{2}}e^{{\cal E}^{*}_{2}\tau}\Big].
\end{eqnarray}
This may be symmetrized so that $B_{1}$ and $B_{2}$ appear equally. When added to
(4.2) it merely introduces the wave function correction $1-B_{j}\approx Z_{j}$.

{\it Cancellation of Unphysical Cuts:}
The $\tau$ dependence in (4.10) produced the unphysical cuts in (4.9). 
When the second line of (5.2) is substituted into (5.1) it gives
\begin{eqnarray}
-g^{2}\int d\Omega_{12} &&
\sum_{\ell=1}^{\infty}e^{i\omega_{\ell}\tau}
{2T\Gamma_{1}\omega_{\ell}
\over (\omega_{\ell}^{2}+{\cal E}_{1}^{2})(\omega_{\ell}^{2}+{\cal E}_{1}^{*2})}
\nonumber\\
\times&&\Big[[1+n({\cal E}_{2})]{1\over 2E_{2}}e^{-{\cal E}_{2}\tau}
+n({\cal E}^{*}_{2}){1\over 2E_{2}}e^{{\cal E}^{*}_{2}\tau}\Big].
\end{eqnarray}
This exactly cancels (4.10) so that the one-loop  unphysical cuts are removed.
Obviously the third and fourth lines of (5.2) will produce new unphysical cuts in the
two-loop self-energy. These  will be canceled by higher loop effects. 

{\it Cut for Three Quasiparticles:} The last term in (5.2) requires integrating
in $k_{0}$ around the branch cuts in the one-loop self energy:
\begin{equation}
i\int_{Im\;k_{0}<0}^{\rm cuts}{dk_{0}\over 2\pi}
{\Pi_{R}^{I}(k)[1+n(k_{0})]\over (k_{0}-{\cal E}_{1})^{2}(k_{0}+{\cal E}_{1}^{*})^{2}}
e^{-k_{0}\tau}.\end{equation}
It is convenient use the representation (4.4) but to change the internal momentum variables 
to $k_{3}$ and $k_{4}$ in correspondence with Fig. 2: 
\begin{displaymath}
\Pi_{R}^{I}(k_{0})
={g^{2}\over 2}\int\! d\Omega_{34}\!
\sum_{\Lambda_{3},\Lambda_{4}}
f(\Lambda_{3})f(\Lambda_{4}){1-e^{-(\Lambda_{3}+\Lambda_{4})\beta}
\over k_{0}-\Lambda_{3}-\Lambda_{4}}.
\end{displaymath}
The denominator $k_{0}-\Lambda_{3}-\Lambda_{4}$ produces the branch cut in $k_{0}$.
The integration around the cut is performed by interchanging the order of integration
to get
\begin{displaymath}
{g^{2}\over 2}\int\! d\Omega_{34}\!
\sum_{\Lambda_{3},\Lambda_{4}}
{f(\Lambda_{3})f(\Lambda_{4})\,e^{-(\Lambda_{3}+\Lambda_{4})\tau}\over
(\Lambda_{3}+\Lambda_{4}-{\cal E}_{1})^{2}
 (\Lambda_{3}+\Lambda_{4}+{\cal E}_{1}^{*})^{2}}. \end{displaymath}
This  is the explicit evaluation of ${\cal D}^{I}_{\rm cut}(\tau,\vec{k})$, i.e. the last line of
(5.2). When substituted into (5.1) the contribution to $\Pi^{II}_{A}(\tau,\vec{k})$ is
\begin{displaymath}
-{g^{4}\over 2}\!\int\! d\Omega_{12}d\Omega_{34}
\sum_{\{\Lambda\}}
{f(\Lambda_{2})f(\Lambda_{3})f(\Lambda_{4})e^{-(\Lambda_{2}+\Lambda_{3}+\Lambda_{4})\tau}\over
(\Lambda_{3}+\Lambda_{4}-{\cal E}_{1})^{2} (\Lambda_{3}+\Lambda_{4}+{\cal
E}_{1}^{*})^{2}}.
\end{displaymath}
This is easily transformed to get $\Pi^{II}_{A}(i\omega_{n},\vec{k})$.
The extension from $i\omega_{n}$ to complex $k_{0}$ analytic in the upper half-plane is
\begin{eqnarray}
\Pi^{II}_{A}(k_{0},\vec{k})=-{g^{4}\over 2}&&\!\int\! d\Omega_{12}d\Omega_{34}
\sum_{\{\Lambda\}}{f(\Lambda_{2})f(\Lambda_{3})f(\Lambda_{4})\over
k_{0}-\Lambda_{2}-\Lambda_{3}-\Lambda_{4}}\nonumber\\
\times&&{e^{-(\Lambda_{2}+\Lambda_{3}+\Lambda_{4})\beta} -1\over
(\Lambda_{3}+\Lambda_{4}-{\cal E}_{1})^{2} (\Lambda_{3}+\Lambda_{4}+{\cal
E}_{1}^{*})^{2}}.
\end{eqnarray}
This contains the cuts for  three quasiparticles at
$k_{0}=\Lambda_{2}+\Lambda_{3}+\Lambda_{4}$.  The unphysical values of $\Lambda$
will be canceled by higher loops. This completes the analysis of Fig. 2.

\subsection{Vertex Correction}

Fig. 3 shows the two-loop diagram containing a vertex correction. Two of the loop momenta are
independent. For definiteness  the independent momenta are taken as  
 $\vec{k}_{1}$ and $\vec{k}_{3}$ and 
$d\Omega\equiv d^{3}k_{1}d^{3}k_{3}/(2\pi)^{3}$.
 The remaining $\vec{k}_{2}, \vec{k}_{4},\vec{k}_{5}$ are linear
combinations of the these two and the external $\vec{k}$. The
self-energy contribution is  
\begin{eqnarray}
\Pi^{II}_{B}(\tau,\vec{k})=
&&{g^{4}\over 4}\int\! d\Omega\int_{0}^{\beta}d\tau'\int_{0}^{\beta}d\tau''\;
{\cal D}_{1}(\tau'){\cal D}_{2}(\tau'')\nonumber\\
\times &&{\cal D}_{3}(\tau''-\tau){\cal D}_{4}(\tau'-\tau)
 {\cal D}_{5}(\tau''-\tau').
\end{eqnarray} 
The three times $\tau, \tau',$ and $\tau''$ 
lie in the interval $[0,\beta]$ and may be ordered in six different ways as follows:
\begin{eqnarray}
B1:\tau'<\tau''<\tau\hskip1cm && B2:\tau''<\tau'<\tau\nonumber\\
B3:\tau<\tau'<\tau''\hskip1cm && B4:\tau<\tau''<\tau'\nonumber\\
B5:\tau'<\tau<\tau''\hskip1cm && B6:\tau''<\tau<\tau'\nonumber\end{eqnarray}
The left and right columns differ by an interchange of  $\tau'$ and $\tau''$.
Because of the structure of the integral, this is the same as 
interchanging $\Lambda_{1}\leftrightarrow
\Lambda_{2}$ and $\Lambda_{3}\leftrightarrow\Lambda_{4}$.
Thus only B1, B3, and B5 need to be computed.
With the representation (3.6) for the quasiparticle propagators, the integration over B1 gives
\begin{eqnarray}
\Pi^{II}_{B1}(\tau,\vec{k})=&&{g^{4}\over 4}\int\! d\Omega
\sum_{\{\Lambda\}}\prod_{j=1}^{5}f(\Lambda_{j})
\nonumber\\
\times\Big[&&{e^{-(\Lambda_{1}+\Lambda_{2})\tau}\over (\Lambda_{1}-\Lambda_{4}-\Lambda_{5})
(\Lambda_{1}+\Lambda_{2}-\Lambda_{3}-\Lambda_{4})}\nonumber\\
+&&{e^{-(\Lambda_{3}+\Lambda_{4})\tau}\over (\Lambda_{2}-\Lambda_{3}+\Lambda_{5})
(\Lambda_{1}+\Lambda_{2}-\Lambda_{3}-\Lambda_{4})}\nonumber\\
+&&{e^{-(\Lambda_{2}+\Lambda_{4}+\Lambda_{5})\tau}\over
(\Lambda_{1}-\Lambda_{4}-\Lambda_{5})(-\Lambda_{2}+\Lambda_{3}-\Lambda_{5})}
\Big].
\end{eqnarray}
The $\tau$ dependence of these three terms will easily lead to two-particle cuts at
$k_{0}=\Lambda_{1}+\Lambda_{2}$, $k_{0}=\Lambda_{3}+\Lambda_{4}$, and a three-particle
cut at $k_{0}=\Lambda_{2}+\Lambda_{4}+\Lambda_{5}$.
The next integration, B2, 
 gives the same answer as (5.8) but with the interchanges $\Lambda_{1}\leftrightarrow
\Lambda_{2}$ and $\Lambda_{3}\leftrightarrow\Lambda_{4}$.

Integration B3 can best be done by using the KMS condition to rewrite it as
\begin{eqnarray}
\Pi^{II}_{B3}&&(\tau,\vec{k})=
{g^{4}\over 4}\int d\Omega\int_{\tau}^{\beta}d\tau'\int_{\tau'}^{\beta}d\tau''\;
{\cal D}_{1}(\tau'){\cal D}_{2}(\tau'')\nonumber\\
\times &&{\cal D}_{3}(\beta+\tau-\tau''){\cal D}_{4}(\beta+\tau-\tau')
 {\cal D}_{5}(\beta+\tau'-\tau'').
\nonumber\end{eqnarray}
The time argument for each of the quasiparticle propagators is positive.
For example, for ${\cal D}_{3}$ the time dependence is $\exp[-\Lambda_{3}(\beta+\tau
-\tau'')]$. The integrand written in this form leads to the most convenient form for the final
answer with  $\Pi(\tau)$ a product of terms of the form
 $\exp(-\Lambda\tau)$ as desired.
Direct integration gives
\begin{eqnarray}
\Pi^{II}_{B3}(\tau,\vec{k})=&&{g^{4}\over 4}\int\! d\Omega
\sum_{\{\Lambda\}}\prod_{j=1}^{5}f(\Lambda_{j})
\nonumber\\
\times\Big[&&{e^{-(\Lambda_{1}+\Lambda_{2})\tau}\;e^{-(\Lambda_{3}+\Lambda_{4}+\Lambda_{5})\beta}
\over(\Lambda_{2}-\Lambda_{3}-\Lambda_{5})
(\Lambda_{1}+\Lambda_{2}-\Lambda_{3}-\Lambda_{4})}\nonumber\\
+&&{e^{-(\Lambda_{3}+\Lambda_{4})\tau}\;e^{-(\Lambda_{1}+\Lambda_{2}+\Lambda_{5})\beta}\over
(\Lambda_{1}-\Lambda_{4}+\Lambda_{5})
(\Lambda_{1}+\Lambda_{2}-\Lambda_{3}-\Lambda_{4})}\nonumber\\
+&&{e^{-(\Lambda_{1}+\Lambda_{3}+\Lambda_{5})\tau}\;e^{-(\Lambda_{2}+\Lambda_{4})\beta}\over
(\Lambda_{1}-\Lambda_{4}-\Lambda_{5})(-\Lambda_{2}+\Lambda_{3}+\Lambda_{5})}
\Big].
\end{eqnarray}
The tau dependence of these terms will again produce two particle cuts at
$k_{0}=\Lambda_{1}+\Lambda_{2}$, $k_{0}=\Lambda_{3}+\Lambda_{4}$, but a different
three-particle cut at $k_{0}=\Lambda_{1}+\Lambda_{3}+\Lambda_{5}$.
Integration B4 requires interchanging $\Lambda_{1}\leftrightarrow
\Lambda_{2}$ and $\Lambda_{3}\leftrightarrow\Lambda_{4}$.

The contribution of B5 is more difficult. First use the KMS condition to write it as
\begin{eqnarray}
\Pi^{II}_{B5}&&(\tau,\vec{k})=
{g^{4}\over 4}\int d\Omega\int_{0}^{\tau}d\tau'\int_{\tau}^{\beta}d\tau''\;
{\cal D}_{1}(\tau'){\cal D}_{2}(\tau'')\nonumber\\
\times &&{\cal D}_{3}(\beta+\tau-\tau''){\cal D}_{4}(\tau-\tau')
 {\cal D}_{5}(\beta+\tau'-\tau'').
\nonumber\end{eqnarray}
The integration gives
\begin{eqnarray}
\Pi^{II}_{B5}(\tau,\vec{k})&&={g^{4}\over 4}\int\! d\Omega
\sum_{\{\Lambda\}}\prod_{j=1}^{5}f(\Lambda_{j})\,C
\nonumber\\
\times\Big[&&e^{-(\Lambda_{1}+\Lambda_{2})\tau}\;e^{-(\Lambda_{3}+\Lambda_{5})\beta}
+e^{-(\Lambda_{3}+\Lambda_{4})\tau}\;e^{-\Lambda_{2}\beta}\nonumber\\
-&&e^{-(\Lambda_{1}+\Lambda_{3}+\Lambda_{5})\tau}\;e^{-\Lambda_{2}\beta}
-e^{(-\Lambda_{2}-\Lambda_{4}+\Lambda_{5})\tau}\;e^{-(\Lambda_{3}+\Lambda_{5})\beta}
\Big]\nonumber\\
C\equiv &&{1\over (\Lambda_{1}-\Lambda_{4}+\Lambda_{5})(-\Lambda_{2}+\Lambda_{3}
+\Lambda_{5})}.\end{eqnarray}
The last term contains tau dependence $\exp(+\Lambda_{5}\tau)$ which, when Fourier
transformed, is difficult to extend analytically in the upper half-plane. It is useful to
isolate all the $\Lambda_{5}$ dependence of this term by defining
\begin{displaymath}
Q\equiv {e^{-\Lambda_{5}(\beta-\tau)}\over
(\Lambda_{1}-\Lambda_{4}+\Lambda_{5})(-\Lambda_{2}+\Lambda_{3}
+\Lambda_{5})}.\end{displaymath}
The generalized KMS relation (C9) proven in Appendix C shows that
\begin{eqnarray}
&&\sum_{\Lambda_{5}}f(\Lambda_{5})Q=\sum_{\Lambda_{5}}
\Big[{f(\Lambda_{5})\,e^{-\Lambda_{5}\tau}\over
(\Lambda_{1}-\Lambda_{4}-\Lambda_{5})(-\Lambda_{2}+\Lambda_{3}-\Lambda_{5})}\nonumber\\
&&+{e^{(-\Lambda_{1}+\Lambda_{4})\tau}F_{5}(-\Lambda_{1}+\Lambda_{4})
-e^{(-\Lambda_{2}+\Lambda_{3})(\beta-\tau)}F_{5}(-\Lambda_{2}+\Lambda_{3})
\over \Lambda_{1}+\Lambda_{2}-\Lambda_{3}-\Lambda_{4}}\Big]
\nonumber\end{eqnarray}
where $F$ is the function defined in (C1).
When this is substituted into (5.10) the result is
\widetext
\begin{eqnarray}
\Pi^{II}_{B5}(\tau,\vec{k})=&&{g^{4}\over 4}\int\! d\Omega
\sum_{\{\Lambda\}}\prod_{j=1}^{4}f(\Lambda_{j})\nonumber\\
\times\Big[&&f(\Lambda_{5}){e^{-(\Lambda_{1}+\Lambda_{2})\tau}\;e^{-(\Lambda_{3}+\Lambda_{5})\beta}
+e^{-(\Lambda_{3}+\Lambda_{4})\tau}\;e^{-\Lambda_{2}\beta}
-e^{-(\Lambda_{1}+\Lambda_{3}+\Lambda_{5})\tau}\;e^{-\Lambda_{2}\beta}
\over (\Lambda_{1}-\Lambda_{4}+\Lambda_{5})(-\Lambda_{2}+\Lambda_{3}+\Lambda_{5})}
\nonumber\\
&&f(\Lambda_{5}){e^{-(\Lambda_{2}+\Lambda_{4}+\Lambda_{5})\tau}\,e^{-\Lambda_{3}\beta}\over
(\Lambda_{1}-\Lambda_{4}-\Lambda_{5})(\Lambda_{2}-\Lambda_{3}+\Lambda_{5})}\nonumber\\
-&&{e^{-(\Lambda_{1}+\Lambda_{2})\tau}\,e^{-\Lambda_{3}\beta}F_{5}(-\Lambda_{1}+\Lambda_{4})
-e^{-(\Lambda_{3}+\Lambda_{4})\tau}\,e^{-\Lambda_{2}\beta}
F_{5}(-\Lambda_{2}+\Lambda_{3})\over 
\Lambda_{1}+\Lambda_{2}-\Lambda_{3}-\Lambda_{4}}\Big].
\end{eqnarray}
The $\tau$ dependence determines the $k_{0}$ dependence.
The terms $\exp(-(\Lambda_{1}+\Lambda_{2})\tau)$ and $\exp(-(\Lambda_{3}+\Lambda_{4})\tau)$
produce two particle cuts at $k_{0}=\Lambda_{1}+\Lambda_{2}$ and $k_{0}=\Lambda_{3}+\Lambda_{4}$.
The terms $\exp(-(\Lambda_{1}+\Lambda_{3}+\Lambda_{5})\tau)$
and $\exp(-(\Lambda_{2}+\Lambda_{4}+\Lambda_{5})\tau)$ produce three particle cuts
at $k_{0}=\Lambda_{1}+\Lambda_{3}+\Lambda_{5}$ and 
$k_{0}=\Lambda_{2}+\Lambda_{4}+\Lambda_{5}$.  Integration B6 requires interchanging
$\Lambda_{1}\leftrightarrow
\Lambda_{2}$ and $\Lambda_{3}\leftrightarrow\Lambda_{4}$.

\narrowtext

\section{Conclusion}

The above results follow from the existence of poles in the full retarded propagator 
$D_{R}'(k_{0},\vec{k}_{j})$ at energies
 $k_{0}=\lambda_{j}$ where
\begin{equation} \lambda_{j}={\cal E}(\vec{k}_{j})\;{\rm or}\; -{\cal
E}^{*}(\vec{k}_{j})\hskip1cm {\rm Im}\lambda<0.
\end{equation}
These  poles were shown to produce  singularities  in  retarded self-energy integrands.
In the 
 two-quasiparticle channels there are singularities at 
$k_{0}=\lambda_{1}+\lambda_{2}$. In the  
three-quasiparticle channels the singularities are at
$k_{0}=\lambda_{1}+\lambda_{2}+\lambda_{3}$. Contributions with
$\lambda={\cal E}$ correspond to stimulated emission of quasiparticles weighted by $1+n({\cal
E})$; contributions with $\lambda=-{\cal E}^{*}$
correspond to absorption of quasiparticles weighted by $n({\cal E})$.

The singularities in the integrands of $\Pi_{R}(k)$ produce branch points when
they are trapped at end points of the three-momentum integrations.   
Without knowing the momentum dependence of ${\cal E}(\vec{k})$ it is only
 possible to analyze this trapping in the equal mass case, i.e. when all the
internal lines have the same dispersion relation ${\cal E}(\vec{k})$.
In that case the pole of the integrand at $k_{0}= {\cal E}(\vec{k}_{1}) +{\cal E}(\vec{k}_{2})$
produces an end point singularity from  $\vec{k}_{1}=\vec{k}_{2}=\vec{k}/2$. The branch
point is thus at $k_{0}=2{\cal E}(\vec{k}/2)$.  For 3 quasiparticles the branch point is at
$k_{0}=3{\cal E}(\vec{k}/3)$.  The poles of the integrand at $k_{0}= {\cal E}(\vec{k}_{1})
-{\cal E}(\vec{k}_{2})^{*}$ and $k_{0}= -{\cal E}(\vec{k}_{1})^{*}
+{\cal E}(\vec{k}_{2})$ produce end point singularities from the region
$\vec{k}_{1}=\alpha\vec{k},\;\vec{k}_{2}=(1-\alpha)\vec{k}$ where $\alpha\to\pm\infty$.
Since all radiative corrections vanish at infinite momentum, the branch points are near the real
axis at $k_{0}=\pm|\vec{k}|-i\eta$. These results hold only for equal masses.  In general the
branch point locations will depend upon the functions ${\cal E}(\vec{k})$.

Cuts in the retarded propagator automatically give those of the advanced propagator because
$D_{A}'(k)=D_{R}'(-k)$. This also determines the four real-time propagators
 \begin{eqnarray}
D'_{11}(k)=&&[1+n(k_{0})]D'_{R}(k)-n(k_{0})D'_{A}(k)\nonumber\\
D'_{12}(k)=&&e^{\sigma k_{0}}n(k_{0})[D'_{R}(k)-D'_{A}(k)]\nonumber\\
D'_{21}(k)=&&e^{(\beta-\sigma)k_{0}}n(k_{0})[D'_{R}(k)-D'_{A}(k)]\nonumber\\
D'_{22}(k)=&&n(k_{0})D'_{R}(k)-[1+n(k_{0})]D'_{A}(k).\end{eqnarray}
Each branch cut of the $D'_{ab}$ is completely below the real axis or completely above.
There are no branch cuts that cross the real axis.
In addition the $D'_{ab}$ have simple
poles at $k_{0}=\pm i2\pi nT$ from the Bose-Einstein functions. Although the
$D'_{ab}$ can be written in terms of the the thermal Feynman propagators $D_{F/\overline{F}}$
this introduces step functions $\theta(k_{0})$ which make the analytic properties of
$D_{F/\overline{F}}$ more complicated.

The real-time self-energies are related to the inverse 
 full propagator by
\begin{equation}
\big[D'(k)\big]^{-1}_{ab}=(k^{2}-m^{2})\sigma^{3}_{ab}-\Pi_{ab}(k).\end{equation}
In terms of the retarded and advanced self-energies this implies
\begin{eqnarray}
\Pi_{11}(k)=&&[1+n(k_{0})]\Pi_{R}(k)-n(k_{0})\Pi_{A}(k)\nonumber\\
\Pi_{12}(k)=&&e^{\sigma k_{0}}n(k_{0})[-\Pi_{R}(k)+\Pi_{A}(k)]\nonumber\\
\Pi_{21}(k)=&&e^{(\beta-\sigma) k_{0}}n(k_{0})[-\Pi_{R}(k)+\Pi_{A}(k)]\nonumber\\
\Pi_{22}(k)=&&n(k_{0})\Pi_{R}(k)-[1+n(k_{0})]\Pi_{A}(k).\end{eqnarray}

 Several interesting points require further investigation.
The separation of free quasiparticle effects was done by rearranging the propagator.
It would be useful to have a operator method for separating the free quasiparticles
from the interactions. Work on this is in progress. A related problem is whether the
discontinuities can be computed directly without having to compute the
entire self-energy as done here. In the perturbative approach, the cutting rules of Kobes and
Semenoff [10] accomplish this. However their derivation also requires using the operator
structure. The physical significance of the discontinuities requires further investigation. Since
the true branch points lie off the real $k_{0}$ axis it is natural that the discontinuities
across the branch cuts are complex. For example, the two-particle discontinuity of (4.5) is
\begin{eqnarray}
{\rm Disc}\,\Pi_{R}(k)=
-i{g^{2}\over 2}&&\int{d\Omega_{12}\over 2E_{1}2E_{2}}
2\pi\delta(k_{0}-{\cal E}_{1}-{\cal E}_{2})\nonumber\\
\times\big[[1+n({\cal E}_{1})]&&[1+n({\cal E}_{2})]-n({\cal E}_{1})n({\cal E}_{2})\big].
\end{eqnarray}
This is very much like what would be expected for the difference between the
production rate of two quasiparticles minus their absorption rate, except that
the quasiparticle energies ${\cal E}$ are complex.

\acknowledgments

This work was supported in part by National Science Foundation grant
 PHY-9630149. It is a pleasure to thank the Institute for Nuclear Theory
at the University of Washington and the Department of Energy for
partial support during the completion of this work.

\appendix

\section{Breakdown of Perturbation Theory}

If one applies the Kobes-Semenoff cutting rules [10] to  Fig. 2 using free thermal propagators it
 has the same breakdown near threshold as the $T=0$ example discussed in Sec 1A.
 The formula for this particular discontinuity is displayed in Le Bellac [14] and in
Gelis [15]. The two-particle discontinuity is
\begin{eqnarray}
{\rm Disc}\,\Pi_{R}(k)=&&{-ig^{2}\over 8\pi^{2}}\int d^{4}p\,
\big[1+n(p_{0})+n(k_{0}-p_{0})\big]\nonumber\\
&&\times\epsilon(p_{0})\delta'(p^{2}-m^{2}){\rm Re}\Pi_{R}(p_{0})\nonumber\\
&&\times\epsilon(k_{0}-p_{0})\delta((p-k)^{2}-m^{2}).
\end{eqnarray}
The contribution of ${\rm Im}\,\Pi_{R}$  has been dropped since it produces a three-particle
discontinuity. To display the result it is useful to let $k=|\vec{k}|$ and $K^{2}=k_{0}^{2}
-\vec{k}^{2}$ and $\alpha=\big(1-4m^{2}/K^{2})^{1/2}$.
Direct integration gives
\begin{eqnarray}
{\rm Disc}\,\Pi_{R}(k)&&={-ig^{2}\over 32\pi
\alpha kK^{2}}\big[1+n({k_{0}+\alpha k\over 2})+n({k_{0}-\alpha k\over 2})\big]\nonumber\\
\times&&\Big[(k+\alpha k_{0}){\rm Re}\Pi_{R}({k_{0}-\alpha k\over 2})
\nonumber\\
&&+(k-\alpha k_{0}){\rm Re}\Pi_{R}({k_{0}+\alpha k\over 2})\Big]
\epsilon(k_{0}-\alpha k)\end{eqnarray}
where kinematics requires that either $K^{2}<0$ or $K^{2}>4m^{2}$.
At the perturbative two-particle threshold, $K^{2}\to 4m^{2}$, so that $\alpha\to 0$ and
\begin{equation}
{\rm Disc}\,\Pi_{R}(k)\to{-ig^{2}\over 16\pi
\alpha K^{2}}\big[1+2n({k_{0}\over 2})\big]{\rm Re}\Pi_{R}({k_{0}\over 2})
\end{equation}
The behavior of this discontinuity like $(1-4m^{2}/K^{2})^{-1/2}$ produces an 
infinite correction at the false threshold which signals the breakdown of perturbation theory
just as in the zero-temperature example of Sec 1A. One can also check from (A2) that
at the lightcone threshold, $K^{2}\to 0^{-}$, the discontinuity does not diverge.
In retrospect, this is because the quasiparticle effects do not change the location of the
space-like branch cut for equal masses,
$-|\vec{k}|<k_{0}<|\vec{k}$, as discussed in Sec 6.

\section{Reality and KMS Conditions}

It is not obvious that the quasiparticle propagator ${\cal D}(\tau,\vec{k})$ displayed in (3.4)
and used  throughout the paper satisfies the  reality and KMS conditions claimed in (3.8) and
(3.9). The infinite sum in (3.4)  obscures these properties.
One can rewrite that sum in  another way using
\begin{displaymath}
e^{i\omega_{n}|\tau|}=e^{-i\omega_{n}|\tau|}+2i\sin(\omega_{n}|\tau|).\end{displaymath}
The sum over $\sin(\omega_{n}|\tau|)$ can be performed using the identity
\begin{eqnarray}
T\sum_{n=1}^{\infty}\sin(\omega_{n}|\tau|)&&{-2i\Gamma\omega_{n}
\over (\omega_{n}^{2}+{\cal E}^{2})(\omega_{n}^{2}+{\cal E}^{*2})}
\nonumber\\
={1\over 4E}\Big[-&&[1+n({\cal E})]e^{-{\cal E}|\tau|}
-n({\cal E}^{*})e^{{\cal E}^{*}|\tau|}\nonumber\\
+&&[1+n({\cal E}^{*})]e^{-{\cal E}^{*}|\tau|}
+n({\cal E})e^{{\cal E}|\tau|}
\Big].\end{eqnarray}
Using this in (3.4) gives
\begin{eqnarray}
{\cal D}(\tau,\vec{k})=&&{1\over 2E}\Big[[1+n({\cal E}^{*})]e^{-{\cal
E}^{*}|\tau|} +n({\cal E})e^{{\cal E}|\tau|}\Big]\nonumber\\
&&-T\sum_{n=1}^{\infty}e^{-i\omega_{n}|\tau|}{2\Gamma\omega_{n}
\over (\omega_{n}^{2}+{\cal E}^{2})(\omega_{n}^{2}+{\cal E}^{*2})}.\end{eqnarray}
Each term on the right hand side is  the complex conjugate of the corresponding term in the
original expression (3.4). Hence
${\cal D}(\tau,\vec{k})$ is real.

To prove that the quasiparticle propagator satisfies the KMS condition 
requires writing the propagator in yet another way. In the original form (3.4) use
\begin{displaymath}
e^{i\omega_{n}|\tau|}=\cos(\omega_{n}|\tau|)+i\sin(\omega_{n}|\tau|).\end{displaymath}
The sum over $\sin(\omega_{n}|\tau|)$  can be performed with the identity (B1)
to give the result
\begin{eqnarray}
{\cal D}(\tau,\vec{k})={1\over 4E}\Big[&&[1+n({\cal E})]e^{-{\cal E}|\tau|}
+n({\cal E}^{*})e^{{\cal E}^{*}|\tau|}\nonumber\\
+&&[1+n({\cal E}^{*})]e^{-{\cal E}^{*}|\tau|}
+n({\cal E})e^{{\cal E}|\tau|}\Big]\nonumber\\
-T\sum_{n=1}^{\infty}&&\cos(\omega_{n}|\tau|){2\Gamma\omega_{n}
\over (\omega_{n}^{2}+{\cal E}^{2})(\omega_{n}^{2}+{\cal E}^{*2})}.\end{eqnarray}
In this form the KMS condition ${\cal D}(\beta-\tau,\vec{k})={\cal D}(\tau,\vec{k})$
is satisfied manifestly.

\section{Generalized KMS Identities}

In Sec 5B it is necessary to use some relations that are generalizations of the KMS
identity. To demonstrate these it is useful to define
\begin{equation}
F(k_{0})={n(k_{0})\over (k_{0}-{\cal E})(k_{0}+{\cal E}^{*})}
-{n(k_{0})\over (k_{0}-{\cal E})(k_{0}+{\cal E}^{*})}.\end{equation}
This satisfies
\begin{equation}
F(-k_{0})=e^{\beta k_{0}}\,f(k_{0}).\end{equation}
$F$ has poles in the lower-half of the complex $k_{0}$ plane
at $k_{0}=\Lambda$ where $\Lambda\in\{ {\cal E}, -{\cal E}^{*}, -i\omega_{n}\}$.
At the poles
\begin{displaymath}
F(k_{0})\to {e^{-\Lambda \beta}\,f(\Lambda)\over k_{0}-\Lambda},\end{displaymath}
where $f(\Lambda)$ are the functions given in (3.7).
It also has poles in the upper half-plane at $k_{0}=-\Lambda$:
\begin{displaymath}
F(k_{0})\to {-f(\Lambda)\over
k_{0}+\Lambda}.\end{displaymath}

{\it KMS Identity:}  Because $F(k_{0})$ vanishes sufficiently rapidly in all
directions of the complex plane as
$|k_{0}|\to\infty$, the contour integral (C3) vanishes when the contour $C$ is a circle of
infinite radius:
\begin{equation}
0=\oint_{C}{dk_{0}\over 2\pi i}\;F(k_{0})\,e^{k_{0}(\beta-\tau)}
\hskip1cm (0\le\tau\le\beta).\end{equation}
The vanishing of the integral implies that the residues of the lower half-plane poles
cancel those of the upper half-plane:
\begin{equation}
\sum_{\Lambda}f(\Lambda)\,e^{-\Lambda(\beta-\tau)}=\sum_{\Lambda}f(\Lambda)\,e^{-\Lambda\tau}.
\end{equation} 
Since the left and right sides of this are the Euclidean propagator (3.6), this just proves the
KMS theorem
\begin{equation}
{\cal D}(\beta-\tau,\vec{k})={\cal D}(\tau,\vec{k}).\end{equation}

{\it Theorem 1:} For $C$ a circular contour at infinity and $x$ any complex number inside the
contour, the following integral vanishes
\begin{equation}
0=\oint_{C}{dk_{0}\over 2\pi i}\;F(k_{0})\,{e^{k_{0}(\beta-\tau)}\over k_{0}-x}
\hskip1cm (0\le\tau\le\beta).\end{equation}
The contribution to the integral of the poles at $k_{0}=\Lambda$, $k_{0}=-\Lambda$,and $k_{0}=x$
must all cancel. This implies
\begin{equation}
\sum_{\Lambda}f(\Lambda){e^{-\Lambda(\beta-\tau)}\over \Lambda+x}
=\sum_{\Lambda}f(\Lambda){e^{-\Lambda\tau}\over -\Lambda+x}
-F(x)\,e^{x(\beta-\tau)}.\end{equation}
This is a generalization of the KMS identity. If the differential operator $(x+d/d\tau)$ is
applied to both sides of (C7) it reduces to (C4).

{\it Theorem 2:} For the same contour as before and $0\le\tau\le\beta$ the integral (C8) vanishes
\begin{equation}
0=\oint_{C}{dk_{0}\over 2\pi i}\;F(k_{0})\,{e^{k_{0}(\beta-\tau)}\over (k_{0}-x)
(k_{0}-y)}. \end{equation}
Evaluating the integral by Cauchy's theorem gives
\begin{eqnarray}
\sum_{\Lambda}f(\Lambda)&&{e^{-\Lambda(\beta-\tau)}\over (\Lambda+x)(\Lambda+y)}
=\sum_{\Lambda}f(\Lambda){e^{-\Lambda\tau}\over (-\Lambda+x)(-\Lambda+y)}\nonumber\\
&&\hskip0.5cm +{F(x)\,e^{x(\beta-\tau)}-F(y)\,e^{y(\beta-\tau)}\over
x-y}.\end{eqnarray}
Applying $(y+d/d\tau)$ th this reproduces (C7). 
This identity is used in rewriting (5.10) in the form (5.11). Obviously these identities could
be genralized to polynomial denominators of any order.

\section{One-loop Calculation in the Real-Time Formalism}

Calculations may also be done directly in the real-time formalism. This Appendix will compute
the one-loop self-energy in the real-time formalism and show that the answer is the
same as obtained rather easily in Sec 4. 
In the quasiparticle approximation the real-time propagators $D_{ab}(k)$ are the linear
combinations (6.2) of  the approximate retarded and advanced quasiparticle propagators
\begin{displaymath}
D_{R}(k)={1\over (k_{0}-{\cal E})(k_{0}+{\cal E}^{*})}
\hskip0.2cm D_{A}(k)={1\over (k_{0}+{\cal E})(k_{0}-{\cal E}^{*})}.
\end{displaymath}
The retarded self-energy that implied by (6.4) is   
\begin{equation}
(e^{\beta k_{0}}+1)\,\Pi_{R}(k)=e^{\beta k_{0}}\,\Pi_{11}(k)-\Pi_{22}(k).\end{equation}
The one-loop contribution has two  propagators with  momenta $k^{\mu}_{1}$ and
$k_{2}^{\mu}$. Integration
 will be over $k_{1}$ with the other defined by
$k_{2}=k_{1}-k$. The necessary one-loop  self-energies are
\widetext
\begin{eqnarray}
\Pi_{11}(k)=&&{ig^{2}\over 2}\int{d^{4}k_{1}\over (2\pi)^{4}}
n(k_{1})n(k_{2})\big[e^{\beta k_{01}}D_{R}(k_{1})-D_{A}(k_{1})\big]
\big[e^{\beta k_{02}}D_{R}(k_{2})-D_{A}(k_{2})\big]\nonumber\\
\Pi_{22}(k)=&&{ig^{2}\over 2}\int{d^{4}k_{1}\over (2\pi)^{4}}
n(k_{1})n(k_{2})\big[D_{R}(k_{1})-e^{\beta k_{01}}D_{A}(k_{1})\big]
\big[D_{R}(k_{2})-e^{\beta k_{02}}D_{A}(k_{2})\big].\end{eqnarray}
When these are substituted into (D1) the term $D_{A}(k_{1})D_{R}(k_{2})$ cancels. 
The remaining three products of the form $D_{}(k_{1})D_{}(k_{2})$ are  multiplied by
combinations of exponentials that cancel one of the Bose-Einstein functions $n(k_{1})$ or
$n(k_{2})$. The result is
\begin{eqnarray}
(e^{\beta k_{0}}+1)\,\Pi_{R}(k)=&&{ig^{2}\over 2}\int{d^{4}k_{1}\over (2\pi)^{4}}
n(k_{1})D_{A}(k_{2})\big[(e^{\beta k_{0}}+1)D_{R}(k_{1})-
(e^{\beta k_{0}}+e^{\beta k_{01}})D_{A}(k_{1})\big]\nonumber\\
+&&{ig^{2}\over 2}\int{d^{4}k_{1}\over (2\pi)^{4}}
n(k_{2})D_{R}(k_{1})\big[-(e^{\beta k_{0}}+1)D_{A}(k_{2})+
(e^{\beta k_{01}}+1)D_{R}(k_{2})\big].
\end{eqnarray}
Note that $D_{R}(k_{1})D_{A}(k_{2})$ appears in both lines.
It is convenient to compute the first integral by closing the $k_{01}$ contour below.  The
poles in the lower half of the $k_{01}$ come from two sources:  $D_{R}(k_{1})$
has quasiparticle poles at $k_{01}={\cal E}_{1}$ and $k_{01}=-{\cal E}_{1}^{*}$; and 
$n(k_{1})$ has poles at $k_{01}=-i\omega_{\ell}$. After  the $k_{01}$ integration is performed,
there is a common  factor $e^{\beta k_{0}}+1$ on the right hand side. 
The  contribution to $\Pi_{R}$ from the first line of (D3) is 
\begin{eqnarray}
&&{g^{2}\over 2}\int {d^{3}k_{1}\over (2\pi)^{3}}
\;{1\over 2E_{1}}\big[{n({\cal E}_{1})\over (k_{0}-{\cal E}_{1}-{\cal E}_{2})
(k_{0}-{\cal E}_{1}+{\cal E}_{2}^{*})}
-{n(-{\cal E}_{1}^{*})\over (k_{0}+{\cal E}_{1}^{*}-{\cal E}_{2})(k_{0}+{\cal E}_{1}^{*}
+{\cal E}_{2}^{*})}\big]\nonumber\\
+&&{g^{2}\over 2}\int {d^{3}k_{1}\over (2\pi)^{3}}\;T\sum_{\ell=1}^{\infty}
{1\over (k_{0}+i\omega_{\ell}-{\cal E}_{2})(k_{0}+i\omega_{\ell}+{\cal E}_{2}^{*})}
\;{2i\omega_{\ell}({\cal E}_{1}^{*}-{\cal E}_{1})\over (\omega_{\ell}^{2}+{\cal E}_{1}^{2})
(\omega_{\ell}^{2}+{\cal E}_{1}^{*2})},
\end{eqnarray}
where $2E_{1}={\cal E}_{1}+{\cal E}_{1}^{*}$.
To compute the integral on the  second  line of (D3) it is convenient to close the
$k_{01}$ contour above. The poles in the upper-half of the $k_{01}$ plane come from
$D_{A}(k_{2})$ (recall $k_{02}=k_{01}-k_{0}$) at $k_{01}=k_{0}+{\cal E}_{2}^{*}$ and
$k_{01}=k_{0}-{\cal E}_{2}$ and from the Bose-Einstein function $n(k_{2})$ at
$k_{01}=k_{0}+i\omega_{\ell}$. The second line of (D3) contributes 
\begin{eqnarray}
&&{g^{2}\over 2}\int {d^{3}k_{1}\over (2\pi)^{3}}\;{1\over 2E_{2}}
\big[{n({\cal E}_{2}^{*})\over (k_{0}+{\cal E}_{2}^{*}-{\cal E}_{1})
(k_{0}+{\cal E}_{2}^{*}+{\cal E}_{1}^{*})}
-{n(-{\cal E}_{2})\over (k_{0}-{\cal E}_{2}-{\cal E}_{1})(k_{0}-{\cal E}_{2}+{\cal E}_{1}^{*})}
\big]\nonumber\\
+&&{g^{2}\over 2}\int {d^{3}k_{1}\over (2\pi)^{3}}
\;T\sum_{\ell=1}^{\infty}{1\over (k_{0}+i\omega_{\ell}-{\cal E}_{1})
(k_{0}+i\omega_{\ell}+{\cal E}_{1}^{*})}{2i\omega_{\ell}({\cal E}_{2}^{*}-{\cal E}_{2})
\over (\omega_{\ell}^{2}+{\cal E}_{2}^{2})(\omega_{\ell}^{2}+{\cal E}_{2}^{*2})}.
\end{eqnarray}
The sum of (D4) and (D5) gives $\Pi_{R}(k)$ to one-loop order. It agrees completely with
the sum of (4.5)-(4.9).  In this method of calculating, the unphysical branch cuts produced by
the denominators containing $k_{0}+i\omega_{\ell}+z$ arise from poles in the Bose-Einstein
functions. They are not artifacts of the Euclidean calculation performed in Sec 4.

\narrowtext

\vfill\eject

\begin{picture}(200,100)
\thicklines
\put(100,50){\oval(100,50)}
\put(0,50){\line(1,0){50}}
\put(150,50){\line(1,0){50}}
\put(50,75){1}
\put(50,20){2}
\end{picture}

Fig 1: One-loop self-energy.

\begin{picture}(200,150)
\thicklines
\put(100,50){\oval(100,50)}
\put(100,75){\oval(60,40)[t]}
\put(0,50){\line(1,0){50}}
\put(150,50){\line(1,0){50}}
\put(45,70){1}
\put(150,70){1}
\put(100,10){2}
\put(100,65){3}
\put(100,85){4}
\end{picture}

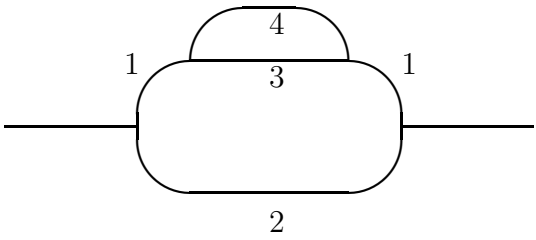
Fig 2: Two-loop self-energy due to one self-energy insertion.

\begin{picture}(200,150)
\thicklines
\put(100,50){\oval(100,50)}
\put(100,25){\line(0,1){50}}
\put(0,50){\line(1,0){50}}
\put(150,50){\line(1,0){50}}
\put(50,75){1}
\put(50,20){2}
\put(140,20){3}
\put(140,75){4}
\put(105,45){5}
\end{picture}

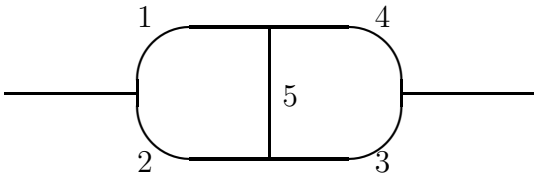
	Fig 3: Two-loop self-energy due to vertex correction.

\end{document}